\documentclass[10pt,twocolumn,letterpaper]{article}

\usepackage{cvpr}              

\usepackage{graphicx}
\usepackage{amsmath}
\usepackage{amssymb}
\usepackage{booktabs}

\usepackage{multirow}
\DeclareMathOperator*{\argmin}{argmin}   

\usepackage[pagebackref,breaklinks,colorlinks]{hyperref}

\usepackage[capitalize]{cleveref}
\crefname{section}{Sec.}{Secs.}
\Crefname{section}{Section}{Sections}
\Crefname{table}{Table}{Tables}
\crefname{table}{Tab.}{Tabs.}

\def\cvprPaperID{3819} 
\def\confName{CVPR}
\def\confYear{2022}

\begin{document}

\title{Learning Optimal K-space Acquisition and Reconstruction using Physics-Informed Neural Networks}

\author{Wei Peng$^{1,3}$ \quad
    Li Feng$^2$ \quad 
    Guoying Zhao$^{1,}$\thanks{Corresponding Author.} \quad 
    Fang Liu$^{3,*}$
    \\
    $^1$CMVS, University of Oulu \quad 
    $^2$Icahn School of Medicine at Mount Sinai \quad 
    $^3$Harvard Medical School
}
\maketitle

\begin{abstract}
   The inherent slow imaging speed of Magnetic Resonance Image (MRI) has spurred the development of various acceleration methods, typically through heuristically undersampling the MRI measurement domain known as k-space. Recently, deep neural networks have been applied to reconstruct undersampled k-space data and have shown improved reconstruction performance. While most of these methods focus on designing novel reconstruction networks or new training strategies for a given undersampling pattern, \textit{e.g.}, Cartesian undersampling or Non-Cartesian sampling, to date, there is limited research aiming to learn and optimize k-space sampling strategies using deep neural networks. This work proposes a novel optimization framework to learn k-space sampling trajectories by considering it as an Ordinary Differential Equation (ODE) problem that can be solved using neural ODE. In particular, the sampling of k-space data is framed as a dynamic system, in which neural ODE is formulated to approximate the 
 system with additional constraints on MRI physics. In addition, we have also demonstrated that trajectory optimization and image reconstruction can be learned collaboratively for improved imaging efficiency and reconstruction performance. Experiments were conducted on different in-vivo datasets (\textit{e.g.}, brain and knee images) acquired with different sequences. Initial results have shown that our proposed method can generate better image quality in accelerated MRI than conventional undersampling schemes in Cartesian and Non-Cartesian acquisitions.
\end{abstract}

\section{Introduction}
\label{sec:intro}
Magnetic Resonance Imaging (MRI) is a powerful clinical tool for disease diagnosis~\cite{li2006modern}. MRI is non-invasive, radiation-free, and can provide excellent soft-tissue contrast, making it an ideal imaging modality for various neurological, oncological, and musculoskeletal applications~\cite{brookeman2004mri}. Standard MRI acquisition sequentially collects the k-space data using segmented sampling patterns composed of multiple shots (\textit{e.g.}, spokes, or interleaves)~\cite{brookeman2004mri,bernstein2004handbook}. However, this acquisition scheme typically requires a long scan time, which is challenging for subjects who cannot tolerate long scans due to injuries, pain, discomfort, or claustrophobia. Notably, the long scan also makes data acquisition sensitive to different types of motion, which can cause image degradation reflected as motion artifact~\cite{zaitsev2015motion}. Therefore, accelerated MRI techniques have played an essential role in reducing MRI acquisition time.

Parallel imaging~\cite{sodickson1997simultaneous,griswold2002generalized,pruessmann1999sense} and compressed sensing~\cite{lustig2007sparse,lustig2010spirit,otazo2010combination} have been extensively investigated for accelerating MRI over the past decades. Parallel imaging uses multi-coil information to estimate the missing k-space data; compressed sensing leverages image sparsity to reconstruct undersampled k-space data using a constrained nonlinear image reconstruction framework. Both techniques can increase imaging efficiency in MRI. Recently, deep learning methods using neural networks have been investigated to reconstruct undersampled k-space data through learning multilevel image representation to remove image artifacts and noises. Various techniques have been proposed using unroll network~\cite{schlemper2017deep,hammernik2018learning,biswas2019dynamic}, end-to-end mapping~\cite{wang2016accelerating,han2018deep,eo2018kiki}, domain transfer learning~\cite{zhu2018image,akccakaya2019scan}, adversarial learning~\cite{mardani2018deep,quan2018compressed}, and unsupervised learning~\cite{liu2021magnetic}. However, while those methods focus on developing novel reconstruction networks or improving network training strategies, very few studies have investigated the optimization of k-space acquisition for learning-based reconstruction. The k-space undersampling patterns are usually kept the same as those previously used in parallel imaging and compressed sensing.

This study will investigate accelerating MRI by learning k-space sampling to maximize image acquisition efficiency for deep learning-based image reconstruction. The k-space acquisition is formulated as a dynamic optimization process and is solved using a neural Ordinary Differential Equation (ODE)~\cite{chen2018neuralode}. This ODE system is first initialized with regular Cartesian and Non-Cartesian k-space trajectories, and the trajectories are then dynamically adjusted towards an optimal pattern that provides the best acquisition efficiency. Furthermore, a joint training strategy to optimize both the neural ODE system and a deep learning reconstruction model is implemented, providing optimal data acquisition and image reconstruction. Experiments were conducted on a publicly available MRI database, fastMRI~\cite{zbontar2018fastmri}, to evaluate the proposed method. A comparative study showed that the proposed method could improve reconstruction performance for Cartesian and Non-Cartesian acquisition using different MRI image sequences. The contributions of this work are summarized as followings:
\begin{itemize}
\vspace{-2mm}
\setlength\itemsep{-0.5em}
    \item The paper introduces a novel k-space acquisition optimization strategy for accelerated MRI using neural ODE, which has never been presented previously to the best of our knowledge. 
    \item We synergistically combined neural ODE and a deep learning-based reconstruction system to optimize k-space trajectories and a reconstruction model towards maximal acquisition efficiency and optimal reconstruction performance for Cartesian and Non-Cartesian acquisitions.
    \item The k-space trajectory optimization is conditioned on MRI hardware constraints considering the gradient amplitude and slew rate limits. This creates a physically feasible trajectory for practical MRI acquisition.
    \item Various experiments demonstrated that the proposed method could significantly improve acquisition efficiency and image quality for Cartesian and Non-Cartesian imaging in different anatomical structures and image contrasts. 
\end{itemize}

\section{Related Works}

K-space undersampling is commonly used in accelerated MRI. Uniform undersampling has been the standard acquisition pattern in parallel imaging~\cite{sodickson1997simultaneous,griswold2002generalized,pruessmann1999sense}. However, conventional parallel imaging methods typically allow for only an acceleration rate of 2-3, and excessive acceleration can lead to severe noise amplification and residual aliasing artifacts. A variable-density random undersampling can create noise-like imaging artifacts and is preferable in modern compressed sensing reconstruction~\cite{lustig2007sparse,lustig2010spirit}. This undersampling strategy typically uses higher sampling density in the low-frequency region of k-space, following the energy distribution in the frequency domain. Although this undersampling scheme achieves image acceleration at multi-dimensional image reconstruction, it remains heuristic and ignores the intrinsic k-space features, which might provide important information for improving reconstruction quality.

A few compressed sensing studies have investigated optimizing k-space acquisition to improve reconstruction performance. Haldar and Kim~\cite{haldar2019oedipus} designed sampling patterns based on constrained Cramer-Rao lower bound with classical experiment design techniques. Sherry \etal~\cite{sherry2020learning} framed the k-space acquisition as a bilevel optimization problem to simultaneously learn the optimal sampling pattern and regularization parameters in image reconstruction. Sanchez \etal~\cite{sanchez2020scalable} learned optimal probability distribution to select a stochastic mask under a given acquisition constraint for dynamic MRI. All those methods are limited to optimize Cartesian sampling under a compressed sensing framework. Recently, Lazarus \etal~\cite{lazarus2019sparkling} proposed an optimization method for accelerated MRI targeting Cartesian and Non-Cartesian k-space acquisition. Their method uses a gradient descent optimizer to minimize the difference between the optimized k-space sampling distribution and a heuristically determined density.

With the rise of deep learning~\cite{lecun2015deep}, a few studies have also proposed deep learning-based approaches for optimizing k-space acquisition. Zhang \etal~\cite{zhang2019reducing} proposed leveraging adversarial learning~\cite{goodfellow2014generative} to select k-space phase-encoding lines to reduce pixel-wise uncertainty on reconstructed MR images. Tim \etal~\cite{bakker2020experimental} formulated optimizing k-space as solving a decision-making problem through a policy search~\cite{david2019RL}. Bahadir \etal~\cite{bahadir2020deep} searched for the optimal sampling patterns by formulating a multivariate Bernoulli distribution on the Cartesian grid through Gumbel-Softmax gradient optimization. Those methods are limited to Cartesian trajectory optimization, and extending them to non-Cartesian imaging could be non-trivial. More recently, Weiss \etal~\cite{weiss2019pilot} proposed using a neural network to learn k-space sampling point locations of Non-Cartesian acquisition. This algorithm is sensitive to selecting initial trajectory points, resulting in a trajectory with a rapidly changing gradient waveform.

Our study proposes a new framework to optimize k-space sampling trajectories by considering it as solving Ordinary Differential Equation (ODE) using neural ODE~\cite{chen2018neuralode}. Neural ODE is a new family of neural networks, which has recently been applied in many applications, including time-series modeling~\cite{grathwohl2018ffjord}, dynamic optimization~\cite{chen2018neuralode}, and image generation~\cite{song2020score}. Our study is the first work to use neural ODE for MRI acquisition optimization to the best of our knowledge.

\section{Methodology}

\begin{figure*}
    \centering
    \includegraphics[width=0.9\textwidth]{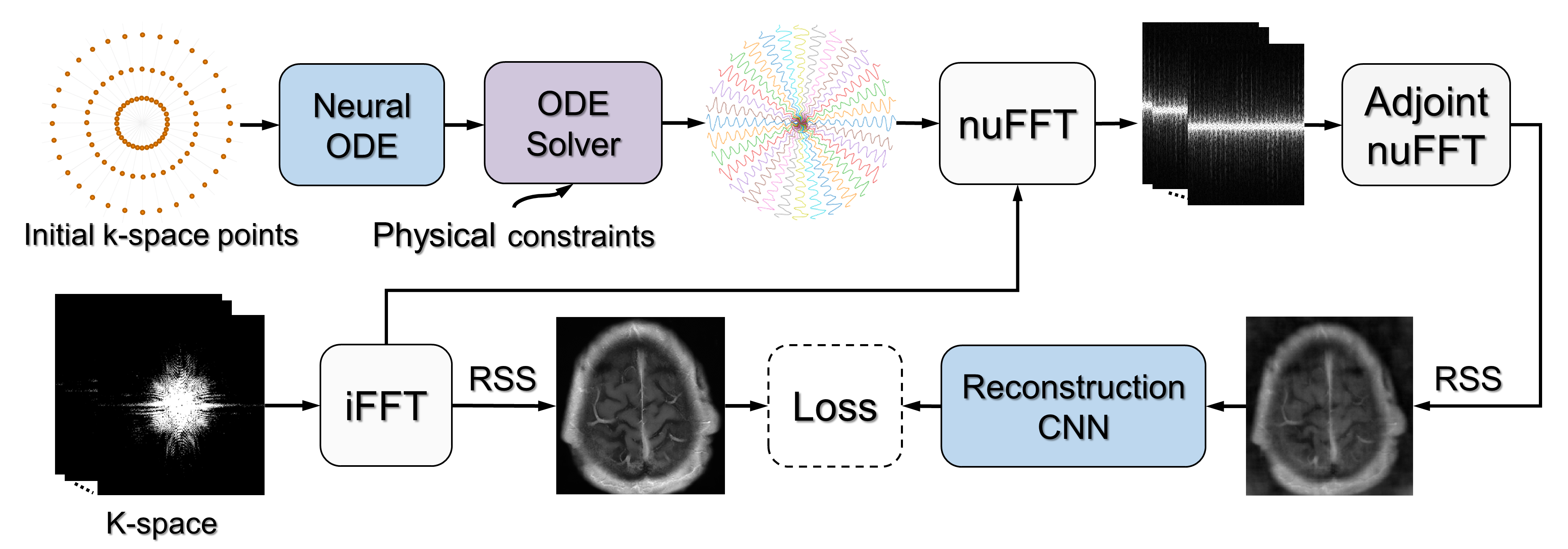}
    \caption{\footnotesize{Schematic illustration of the proposed framework. Given a group of initial k-space control points (yellow points on each radial spoke), a neural ODE is used to approximate the dynamics of the trajectory.Then, a nuFFT is employed to transform the k-space data to the image domain, followed by a root-sum-of-squares reconstruction (RSS) to combine multi-channel images and a reconstruction model to remove artifacts, noises and improve the overall quality of the MR image.}}
    \label{fig:frame}
\end{figure*}

As illustrated in Figure~\ref{fig:frame}, the proposed algorithm consists of two deep learning networks with their associated data processing pathways. First, a neural ODE is introduced to approximate the dynamic evolution of the k-space trajectory from its initial status. With the optimized trajectory, the multi-coil k-space data is undersampled using a non-uniform Fast Fourier Transform (nuFFT). Then, the undersampled data is transferred to image domain. On top of this, a second reconstruction network is trained to remove MR image artifacts and noise. In the following subsections, we provide more details for each component. 

\subsection{Problem Formulation}
ODE mathematically describes variable changes using integrals and derivatives, which can model a dynamic system. Neural ODE~\cite{chen2018neuralode,mcnamara2004fluid} uses a neural network to parameterize the dynamics of the variable. Neural ODE has recently shown impressive performance in sequential data optimization for regular and irregular time sampling~\cite{kidger2020neural}, making it an ideal new approach for optimizing MRI k-space trajectory. In MRI, a k-space trajectory can be described as $k(t)$. Because the k-space trajectory is typically differentiable, the description of the trajectory dynamics can be formulated as an ODE as
\begin{equation}\label{eq:lODE}
    \frac{\mathrm{d} k(t) }{\mathrm{d} t} = f_{\theta}(k(t), t ),
\end{equation}
where $f_{\theta}$ is a neural network with parameters $\theta$. This linear ODE can be simply solved using integration as 
\begin{equation}\label{eq:ODE}
    k(t) = k(0) + \int_{0}^{t} f_{\theta}(k(\tau),\tau) \mathrm{d} \tau,
\end{equation}
where $k(0)$ is an initial trajectory point that can be provided using either Cartesian or Non-Cartesian trajectories (\textit{e.g.}, radial and spiral). We can use $\text{ODEsolver}()$ to represent an ODE solver that takes the inputs of neural network mapping function $f_{\theta}$, initial state (initial k-space control points) $k(0)$,  and time sequence $t$, and then outputs the predicted k-space trajectory 
\begin{equation}\label{eq:ODEsolver}
    k(t) = \text{ODEsolver}(f_{\theta}, k(0), t).
\end{equation}
Suppose $\hat{k}$ is an optimal k-space trajectory that provides the most efficient k-space acquisition. The optimization can then be framed as
\begin{equation}\label{loss1}
    \argmin_{\theta} ~\ell(k(t),\hat{k}) = \argmin_{\theta} ~\ell(\text{ODEsolver}(f_{\theta}, k(0), t), \hat{k}),
\end{equation}
where the loss function $~\ell(k(t),\hat{k})$ minimizes the difference between predicted trajectory $k(t)$ and $\hat{k}$ to update $f_{\theta}$.

In addition, a physically applicable trajectory must obey the hardware constraints of an MRI system,  including the maximum imaging gradient $G_{\text{max}}$, determined by the peak-current, and the maximum gradient slew rate $S_{\text{max}}$, describing the maximum change of gradient in a unit time. The hardware constraints on the maximum gradient and slew rate determine the trajectory speed and acceleration, which can be defined as
\begin{equation}\label{eq:va}
    v =  \frac{\mathrm{d} k(t)}{\mathrm{d} t},
    a =  \frac{\mathrm{d} k(t)}{\mathrm{d} t^2}.
\end{equation}
With these definitions, the k-space trajectory optimization conditioned on MRI physics can be reframed as
\begin{equation}\label{eq:condition}
\begin{split}
    \argmin_{\theta}  & ~\ell(\text{ODEsolver}(f_{\theta}, k(0), t), \hat{k}), \\
\text{subject to}\\
        & v \leq \gamma * G_{\text{max}}\\
        & a \leq \gamma * S_{\text{max}}.
\end{split}
\end{equation}
where the $\gamma$ is the gyromagnetic ratio for a specific nucleus (\textit{e.g.}, proton). To simplify the expression for $\ell(k(t),\hat{k})$, we use $\ell(k(t))$ in the following manuscript.

\subsection{Neural ODE Solver}

The integration operation in the ODE model makes it difficult to efficiently update $f_{\theta}$ using the standard backpropagation approach.  Computing the gradients for all k-space points in the trajectory during the backpropagation will lead to a gigantic computational graph,  which cannot be stored in the memory of typical computing hardware. For example, the loss for a k-space point at the end of the trajectory $t_e$ can be defined as,
\begin{equation}\label{eq:solveODE}
    \ell(k(t_e)) = \ell( k(0) + \int_{0}^{t_e} f_{\theta}(k(\tau),\tau) \mathrm{d} \tau).
\end{equation}
To  back propagate the gradients from $t_e$ to the initial point,  the values of $[\frac{\partial \ell}{\partial k(t) },   \frac{\partial \ell}{\partial t },\frac{\partial \ell}{\partial \theta }]$ at each time point between 0 and $t_e$ in the entire trajectory needs to be computed separately using traditional back propagation approach~\cite{rumelhart1986learning}, which is memory prohibited. In this current study, we tailored a memory-efficient adjoint method~\cite{mcnamara2004fluid} to  optimize  our ODE model in Eq.~(\ref{eq:condition}). Unlike the traditional back propagation approach, the adjont method will compute the dynamics of $[\frac{\partial \ell}{\partial k(t) },   \frac{\partial \ell}{\partial t },\frac{\partial \ell}{\partial \theta }]$, so that the parameter update can be implemented using an integration operation for any time point in the trajectory. For example, the adjoint method will compute the adjoint state $\mathbf{a}_{k}(t)$ for the trajectory $k(t)$, defined as
\begin{equation}
    \mathbf{a}_{k}(t) =  \frac{\partial \ell}{\partial k(t) }.
\end{equation}
Using continuous chain-rule, $\mathbf{a}_{k}(t)$ can be also denoted as
\begin{equation}
    \mathbf{a}_{k}(t) =\frac{\partial \ell}{\partial k(t+\epsilon) } \frac{\partial k(t+\epsilon)}{\partial k(t) }= \mathbf{a}_{k}(t+\epsilon)\frac{\partial k(t+\epsilon)}{\partial k(t) },
\end{equation}
where $\epsilon $ is very small value. The dynamics of the adjoint state ($\frac{\mathrm{d} \mathbf{a}_{k}(t)}{\mathrm{d}t}$) can be calculated  as 
\begin{equation}\label{eq:dynamicA}
\begin{split}
    \frac{\mathrm{d} \mathbf{a}_{k}(t)}{\mathrm{d}t} &=  \lim_{\epsilon \rightarrow 0^+}\frac{\mathbf{a}_{k}(t+\epsilon)- \mathbf{a}_{k}(t)}{\epsilon }\\
                               &=  \lim_{\epsilon \rightarrow 0^+}\frac{-\mathbf{a}_{k}(t+\epsilon) \frac{\partial (\int_{t}^{t+\epsilon} f_{\theta}(k(\tau),\tau) \mathrm{d} \tau)}{\partial k(t) }}{\epsilon }\\
                               &=  - \mathbf{a}_{k}(t)^T \frac{\partial f_{\theta}(k(t), t)}{\partial k(t) },
\end{split}
\end{equation}
Likewise, similar to Eq.~(\ref{eq:dynamicA}), the dynamics of the adjoint $ \mathbf{a}_t(t)$ and $ \mathbf{a}_{\theta}(t)$ can be computed as
\begin{equation}\label{eq:at-diff}
   \frac{\mathrm{d} \mathbf{a}_t(t)}{\mathrm{d} t} = - \mathbf{a}_{k}(t)^T \frac{\partial f_{\theta}(k(t), t)}{\partial t },
\end{equation}

\begin{equation}\label{eq:a-theta-diff}
 \frac{\mathrm{d} \mathbf{a}_{\theta}(t)}{\mathrm{d}t} = - \mathbf{a}_{k}(t)^T \frac{\partial f_{\theta}(k(t), t)}{\partial \theta }.
\end{equation}

Therefore, an $\text{ODEsolver}()$ can be applied to efficiently compute $ \mathbf{a}_k(t)$, $ \mathbf{a}_t(t)$ and $ \mathbf{a}_{\theta}(t)$ to update the corresponding variables $[\frac{\partial \ell}{\partial k(t) },   \frac{\partial \ell}{\partial t },\frac{\partial \ell}{\partial \theta }]$ for all time points between 0 and $t_e$. For example, updating $\theta$ can be computed as $\theta = \theta - \eta \mathbf{a}_{\theta}(t)$, with small $\eta$ determined by the learning rate. Compared with the traditional backpropagation approach, the adjoint method has several advantages: linear scalability with problem size, significantly reduced memory cost, and precise numerical error control. Besides, as this approach is a continuous-time analogy to the traditional backpropagation, it can be easily combined with other gradient-based optimization methods in network training. Therefore, the adjoint method has been recently widely applied in high-performance computation and optimization applications involving integration processes~\cite{chen2018neuralode}. 

To implement the adjoint method in our algorithm, we applied the standard automatic differentiation function in PyTorch~\cite{Paszke2019PyTorchAI} to compute the vector-Jacobian products $\mathbf{a}_{k}(t)^T \frac{\partial f}{\partial k}$,  $\mathbf{a}_{k}(t)^T \frac{\partial f}{\partial t}$, and $\mathbf{a}_{k}(t)^T \frac{\partial f}{\partial \theta}$ in Eqs. (\ref{eq:dynamicA}), (\ref{eq:at-diff}), and (\ref{eq:a-theta-diff}), respectively, as the initial inputs for the $\text{ODEsolver}()$. To further improve the training efficiency, an augmented state combining $k(t)$ and these three adjoint states is used so that a single function call to an $\text{ODEsolver}()$ can compute all integration operations. More specifically, the augmented initial state at time $t_e$ (reversed time order) can be built using a concatenated vector as $s_0 = [k(t_e),\mathbf{a}_{k}(t_e),0_{\theta},  \frac{\partial \ell}{\partial t_e} ]$. As a result, based on Eqs. (\ref{eq:dynamicA}), (\ref{eq:at-diff}), and (\ref{eq:a-theta-diff}), the dynamics of the augmented state denoted as $d(s_0,t, \theta)$, can be defined as 
\begin{equation}
    \begin{split}
        d(s_0,t, \theta) 
        = [f_{\theta}&(k(t),t), -\mathbf{a}_{k}(t)^T \frac{\partial f_{\theta}}{\partial k(t) },\\
        &-\mathbf{a}_{k}(t)^T \frac{\partial f_{\theta}}{\partial \theta },  -\mathbf{a}_{k}(t)^T \frac{\partial f_{\theta}}{\partial t }].
    \end{split}
\end{equation}
Therefore, the gradients at each time point in the trajectory can be computed by employing a single $\text{ODEsolver}()$ call, as
\begin{equation}\label{eq:gradient}
    [k(t), \frac{\partial \ell}{\partial k(t) },\frac{\partial \ell}{\partial \theta },   \frac{\partial \ell}{\partial t }]=\text{ODEsolver}(d(s_0,t,\theta),s_0, t).
\end{equation}
Finally, the trajectory $k(t)$ can be predicted and all gradients are efficiently computed for updating network model.

\subsection{Joint Optimization}
In practice, the optimal k-space trajectory used as training reference is typically unavailable before network training. Therefore, the k-space optimization problem in Eq.~(\ref{eq:condition}) cannot be directly implemented using supervised learning in the k-space domain. Instead, we propose to combine trajectory optimization and image reconstruction into a unified learning framework. As illustrated in Figure~\ref{fig:frame}, an image reconstruction network is utilized to reconstruct k-space data derived from the optimized trajectory. The reconstructed images can then be compared with the fully sampled MR images as the training reference. Mathematically, the optimization in Eq.~(\ref{eq:condition}) can be reformatted as
\begin{equation}\label{eq:loss}
\begin{split}
    \argmin_{\theta,\beta}  &~\ell (h_{\beta}( \Im (\text{ODEsolver}(f_{\theta}, k(0), t), \mathit{I})), \mathit{I}) \\
    & + \lambda_1 ~\text{S}( v, G_{\text{max}} )
      + \lambda_2 ~\text{S}( a,S_{\text{max}}),
\end{split}
\end{equation}
where $\Im$ represents the image encoding operation for generating intermediate images using the k-space trajectories and data through non-uniform Fast Fourier Transform (nuFFT) and adjoint nuFFT~\cite{potts2001fast,fessler2007nuFFT}.  $h_{\beta}$ is an end-to-end mapping network, taking intermediate images as the input and removing residual artifacts and noises. In our study, the loss $ \ell$ uses a hybrid loss function combining both $l$1 loss and Structural Similarity Index Measurement (SSIM) loss~\cite{wang2004image} to characterize both the pixel-wise and structural similarity between the output of the mapping network and the fully sampled MR images $\mathit{I}$.  In addition, because the physical constraints to regularize k-space trajectory are undifferentiable in network training, an element-wise soft shrinkage loss function~\cite{lu2018shrinkage} \text{S} is introduced. More specifically, $\text{S}(b, c)$ returns zero when input $b$ is in the range of $[-c, c]$, otherwise returns the difference between $b$ and $c$. This loss function penalizes the training to ensure the imaging gradient and the gradient slew rate can converge into the maximum allowed values. $\lambda_1$ and $\lambda_2$ are weighting factors to balance the loss terms in this optimization function.

\section{Experiments}
\subsection{In-vivo Image Datasets}
Experiments were conducted to evaluate the proposed method using the image datasets from fastMRI database~\cite{zbontar2018fastmri}, which contains knee and brain MR images acquired with different MR sequences. For the knee images, a subset of knee data on 80 subjects was randomly selected for training (N=60), validation (N=5), and testing (N=15). The knee image dataset was acquired using a 2D coronal proton density-weighted fast spin-echo sequence with the following imaging parameters: echo train length 4, matrix size 320×320, in-plane resolution 0.5mm×0.5mm, slice thickness 3mm, repetition time (TR) ranging between 2200 and 3000ms, and echo time (TE) ranging between 27 and 34ms. The k-space data were fully sampled using a 16-channel knee coil array at 1.5T and 3.0T MR scanners. For the brain images, a subset of the brain data on 65 subjects was randomly selected for training (N=50), validation (N=5), and testing (N=10). The brain datasets include 2D axial T1-weighted images without (AXT1) and with (AXT1POST) the administration of contrast agent. These fully sampled brain images were acquired using multi-channel brain coil arrays at 1.5T and 3.0T MR scanners based on the standard clinical protocol. Because the image size varies across different subjects, the images were unified into a 256x256 matrix size via cropping the central region of images that were obtained by taking the inverse Fourier Transform of the fully sampled k-space data. 
\begin{figure}
    \centering
    \includegraphics[width = 0.38 \textwidth]{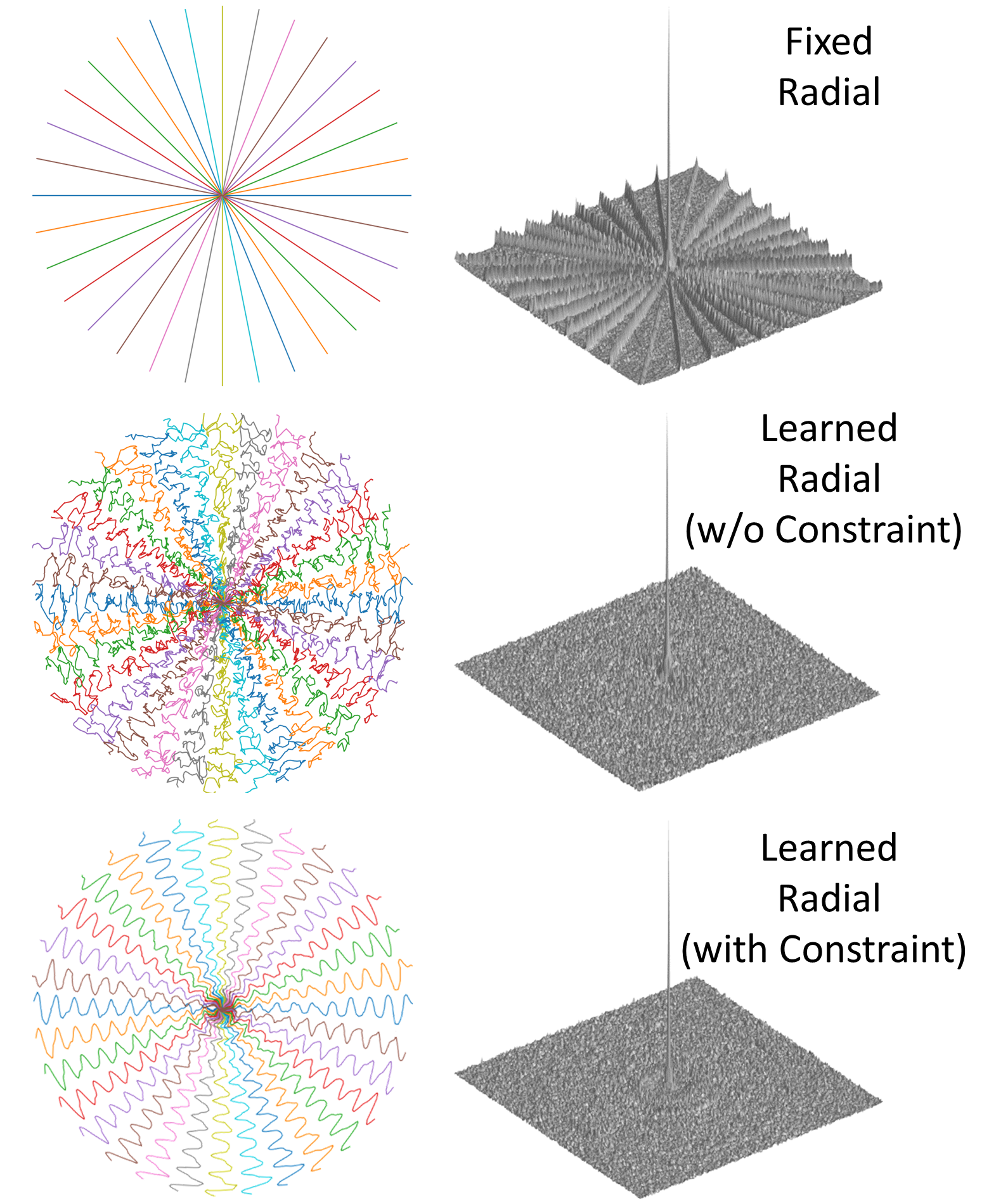}
    \caption{\footnotesize{Trajectory comparison between fixed and learned radial trajectory with 16 spokes, optimized on the brain AXT1POST datasets. \textbf{Top}: initial radial trajectory. \textbf{Middle}: learned radial trajectory without physical constraints. \textbf{Bottom}: learned radial trajectory with physical constraints. The corresponding PSF is shown for each trajectory.}}
    \label{fig:radial_trj}
\end{figure}
To demonstrate the capability of our proposed method for optimizing both Cartesian and Non-Cartesian trajectories, we investigated initializing the trajectory optimization using three different k-space sampling patterns, including Cartesian, spiral(in Appendix), and radial, respectively. Acceleration was implemented using a reduced number of shots, including $N$= 4, 8, 16, and 32, respectively. For the radial initialization, a spoke at 0 degrees was first created, and then the other $N-1$ spokes were built by linearly rotating $\pi/N$ degrees. Each shot was designed to contain 1000 sampling points corresponding to a readout time of 4ms at a dwell time of 4$\mu$s. The maximum imaging gradient was 50mT$/$m, and the maximum slew rate was 200T$/$m$/$s. To reduce the length of the integral in the ODE, each shot was further divided into 100 segments with equal length, and one control point in each segment (resulting in a total of 100 control points per shot) was used as the initial state to predict the remaining 900 sampling points using the described ODE solver in Eq.~(\ref{eq:ODEsolver}). 

Our proposed models were designed using PyTorch\cite{Paszke2019PyTorchAI}. More architecture details are provided in appendix. The ODE solver was implemented using TorchDiffeq~\cite{chen2018neuralode} package, from which a Runge-Kutta of order five of Dormand-Prince-Shampine solver~\cite{calvo1990fifth} was applied given its good computing performance, accuracy, and efficiency in our initial experiment. Because of the GPU memory limit, the multi-channel images were combined into one image using a root-sum-of-squares reconstruction (RSS)~\cite{roemer1990nmr} and then used as the input of the U-Net\cite{ronneberger2015u} reconstruction network. 

The Adam optimizer~\cite{kingma2014adam} was applied to train our networks, and the learning rates were set to 0.01 and 0.001 for the neural ODE and the U-Net, respectively. The framework was trained for a total of 100 epochs. To facilitate the neural ODE convergence, the networks were optimized without the constraints at the first 25 epochs. Then, physical constraints were added to fine-tune the learned trajectory by empirically setting $\lambda_1$ and $\lambda_2$ to $0.1$ for Eq.~(\ref{eq:loss}). All the training and testing were performed using an NVIDIA Tesla K80 GPU with 11GB GDDR5 RAM.

\subsection{Evaluation of Reconstruction}

Image reconstruction was evaluated qualitatively and quantitatively. Quantitative metrics calculate the difference between the reconstructed and fully sampled reference images in the testing datasets. Two metrics were used, focusing on different aspects of the reconstruction quality. The peak signal-to-noise ratio (PSNR) was used to assess the overall reconstructed image errors. The SSIM was used to estimate the overall image similarity to the reference images. A paired nonparametric Wilcoxon signed‐rank test with statistical significance defined as a $p < 0.05$ was used to evaluate the group-wise differences between methods.
\vspace{-3mm}
\subsubsection{Analysis, Visualization, and Evaluation}
\begin{figure}
    \centering
    \includegraphics[width = 0.4 \textwidth]{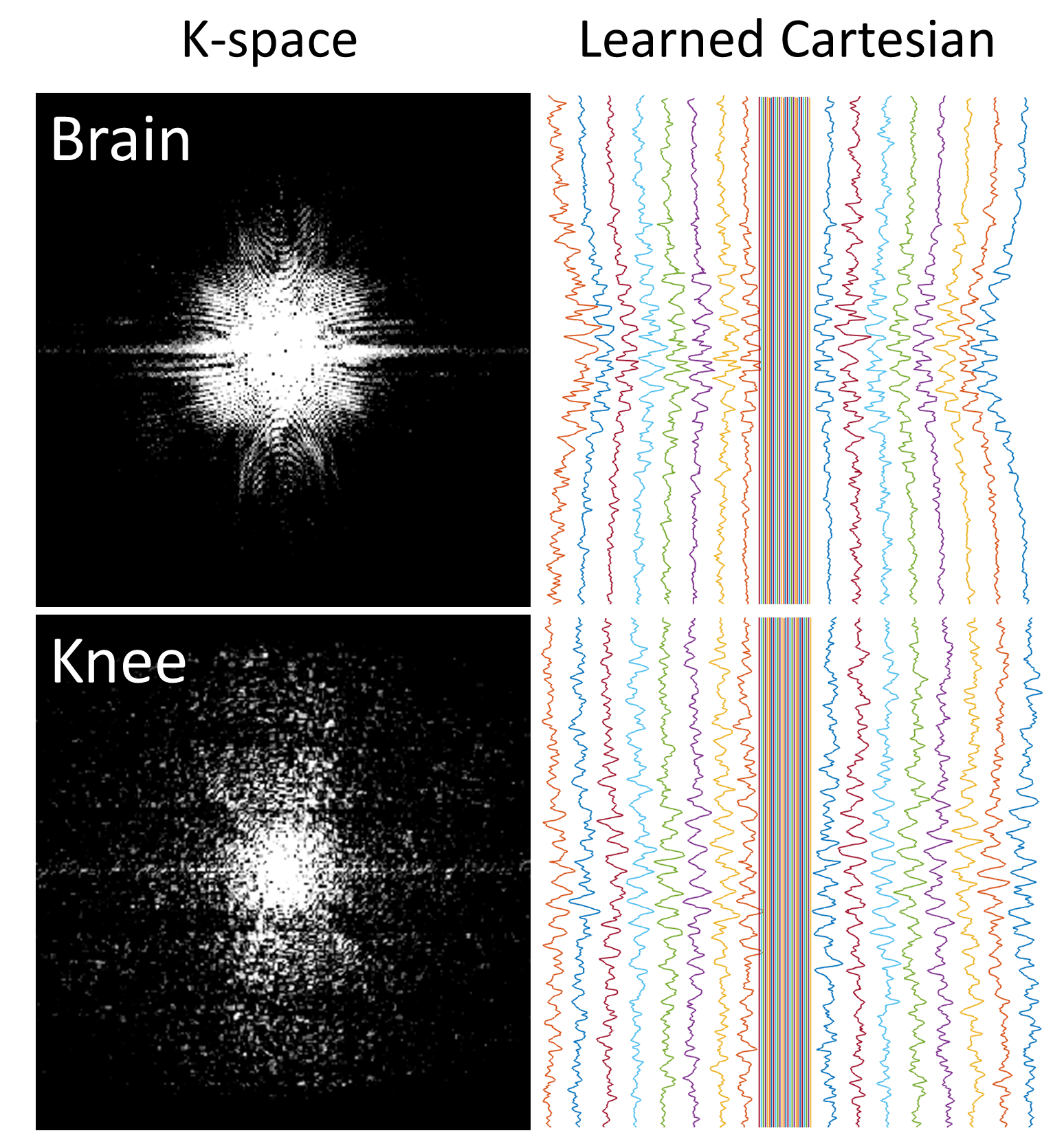}
    \caption{\footnotesize{Demonstration of trajectory optimization difference between different image datasets.} }
    \label{fig:diff-dataset}
    \vspace{0mm}
\end{figure}
\begin{figure*}
    \centering
    \begin{tabular}{c@{\hskip -0.8mm}c@{\hskip -0.8mm}c@{\hskip -0.6mm}c@{\hskip -1mm}c}
       \includegraphics[width = 0.1935\textwidth]{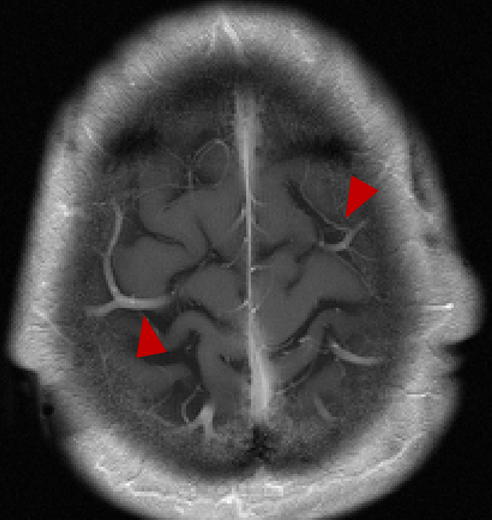}  & \includegraphics[width = 0.2\textwidth]{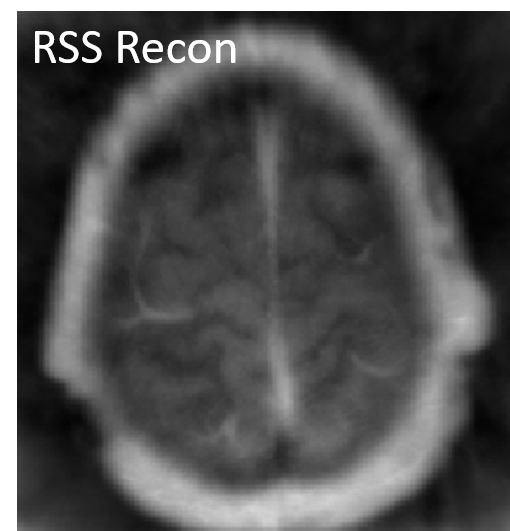}  & \includegraphics[width = 0.2\textwidth]{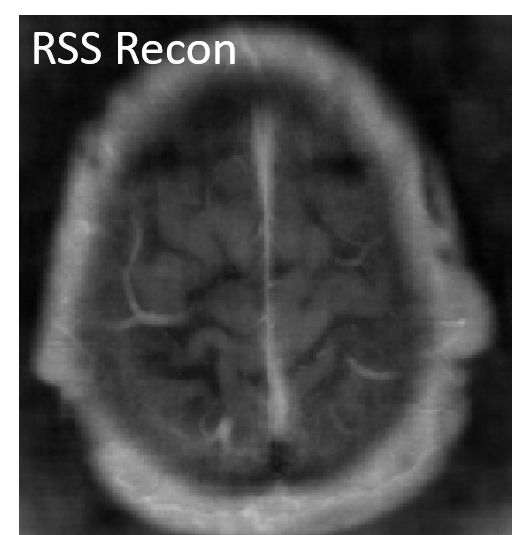}  & \includegraphics[width = 0.2\textwidth]{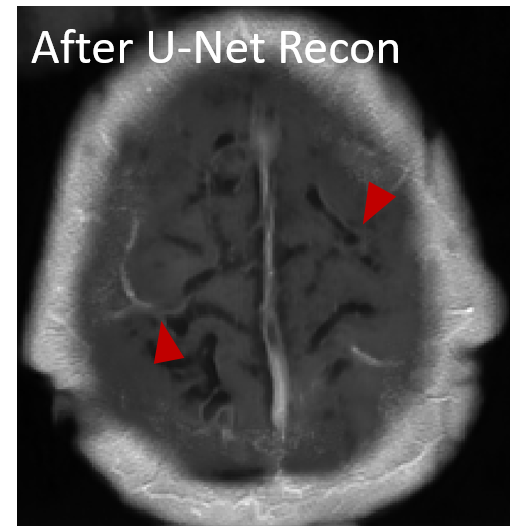}  & \includegraphics[width = 0.2\textwidth]{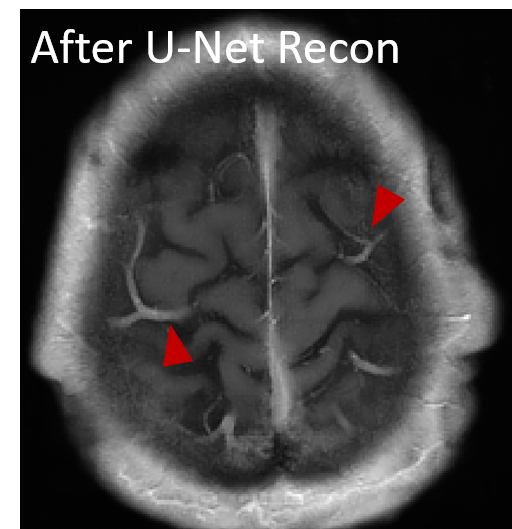}  \\
        Reference & Fixed  & Learned & Fixed & Learned \\
    \end{tabular}
    \caption{\footnotesize{MRI reconstruction comparison between fixed (Fixed) and learned (Learned) radial trajectory with 32 spokes for brain AXT1POST images. The {leftmost column} shows the ground truth fully sampled image. Compared to this, the learned trajectory provides accurate image contract recovery and detailed blood vessel reconstruction superior to the fixed trajectory both before and after the reconstruction, as indicated by the red arrows.}}
    \label{fig:POST-ALL}
\end{figure*}


\begin{table*}
  \caption{ \small{Quantitative comparison on brain AXT1 images at different acceleration levels (Acc. Level).}}
  \label{TableOne}
  \centering
  \scalebox{0.6}{\begin{tabular}{r|cccc|cccc|cccc}
    \toprule
    \multirow{3}{*}{Acc. Level} &\multicolumn{4}{c}{Cartesian} & \multicolumn{4}{c}{Radial}  &    \multicolumn{4}{c}{Spiral}              \\
    \cline{2-13}
    &\multicolumn{2}{c}{Fixed} & \multicolumn{2}{c}{Learned}  & \multicolumn{2}{c}{Fixed} & \multicolumn{2}{c}{Learned}  &\multicolumn{2}{c}{Fixed} & \multicolumn{2}{c}{Learned}               \\
     \cline{2-13}
    & PSNR &SSIM &PSNR &SSIM &PSNR &SSIM &PSNR &SSIM &PSNR &SSIM &PSNR &SSIM  \\
    
   \midrule  
 4-Shots & $29.75 \pm 0.78$  & $0.81 \pm 0.02$ & $\mathbf{30.11 \pm 0.81}$ & $\mathbf{0.82\pm 0.02}$ & $24.24 \pm 0.83$ & $0.65 \pm 0.02$ &$\mathbf{24.98 \pm 0.85}$ & $\mathbf{0.67 \pm 0.02 }$ &$26.94 \pm 0.72 $& $0.74 \pm 0.01$ & $ \mathbf{29.36\pm 0.67} $& $\mathbf{ 0.80 \pm 0.01 }$ \\
    \midrule
8-Shots & $30.54 \pm 0.78$  & $ 0.82 \pm 0.01$ & $ \mathbf{30.55\pm 0.75}$ & $ \mathbf{0.83 \pm 0.01}$ & $ 25.71 \pm 0.89$ & $0.69 \pm 0.02$ &$ \mathbf{28.98 \pm 0.58 } $ & $\mathbf{0.78 \pm 0.01}$ & $29.44\pm 0.98$& $  0.81 \pm 0.01$ &$\mathbf{30.99 \pm 0.64}$ & $\mathbf{0.83\pm 0.01}$\\
    \midrule
16-Shots & $ 31.06 \pm 0.85$ & $0.83  \pm 0.02$ & $\mathbf{31.27 \pm 0.67}$ & {$\mathbf{0.84 \pm 0.01}$} &  $28.38 \pm 1.69 $ & $0.74 \pm 0.02$ & $\mathbf{31.03 \pm 0.59 }$ & $\mathbf{0.82 \pm 0.01}$& $30.62 \pm 1.15$ & $0.84 \pm 0.01$ & $\mathbf{31.73 \pm 0.62}$ & $\mathbf{0.85 \pm 0.01}$ \\
    \midrule
32-Shots &$31.56\pm 0.86$  & $0.84 \pm 0.01$ & $\mathbf{31.75\pm 0.70}$ & $\mathbf{0.85 \pm 0.02}$ & $30.18 \pm 0.73$ & $0.78 \pm 0.01$ & $\mathbf{33.06 \pm 1.23}$ & $\mathbf{0.86 \pm 0.01}$& $32.08 \pm 0.62$ & $ 0.86 \pm 0.01$  & $\mathbf{32.79 \pm 0.65}$ & $\mathbf{0.87 \pm 0.01}$\\
    \bottomrule
  \end{tabular}}
  \vspace{-3mm}
\end{table*}

\begin{table*}
  \caption{\small{Quantitative comparison on brain AXT1POST images at different acceleration levels (Acc. Level).}}
  \label{TableTwo}
  \centering
  \scalebox{0.6}{\begin{tabular}{r|cccc|cccc|cccc}
    \toprule
    \multirow{3}{*}{Acc. Level} &\multicolumn{4}{c}{Cartesian} & \multicolumn{4}{c}{Radial}  &    \multicolumn{4}{c}{Spiral}              \\
    \cline{2-13}
    &\multicolumn{2}{c}{Fixed} & \multicolumn{2}{c}{Learned}  & \multicolumn{2}{c}{Fixed} & \multicolumn{2}{c}{Learned}  &\multicolumn{2}{c}{Fixed} & \multicolumn{2}{c}{Learned}               \\
     \cline{2-13}
    & PSNR &SSIM &PSNR &SSIM &PSNR &SSIM &PSNR &SSIM &PSNR &SSIM &PSNR &SSIM  \\
    
    \midrule 
   4-Shots & $ 29.34 \pm 1.63$  & $0.81 \pm 0.02$ & $\mathbf{30.27 \pm 0.73}$ & $\mathbf{0.83 \pm 0.02}$   & $24.77 \pm 1.36$ & $ 0.66 \pm 0.02 $ & $\mathbf{26.68 \pm 1.09}$ & $\mathbf{0.72 \pm 0.02}$ &$24.96\pm 0.82 $& $0.67 \pm 0.01$ & $\mathbf{29.19 \pm 1.29}$& $\mathbf{0.78 \pm 0.02}$ \\
    \midrule
    8-Shots & $29.54  \pm 1.57$  & $0.81 \pm 0.03$ & $\mathbf{31.00 \pm 0.81}$ & $\mathbf{0.84 \pm 0.02}$  & $ 26.24 \pm 1.36$ & $0.70 \pm 0.02$ &$\mathbf{29.44 \pm 1.05}$ & $\mathbf{0.79 \pm 0.01}$ &$28.34 \pm 1.08 $ & $  0.80\pm 0.02$ &$\mathbf{30.85\pm 0.94}$ & $\mathbf{0.82\pm 0.01}$\\
    \midrule
    16-Shots & $30.12 \pm 1.61 $ &  $0.82 \pm 0.12 $& $\mathbf{31.72 \pm 1.31}$ & $\mathbf{0.85 \pm 0.02}$ &  $28.83 \pm 1.49 $ & $0.77 \pm 0.03$ &$\mathbf{30.78 \pm 1.28}$ & $\mathbf{0.81 \pm 0.03}$& $ 30.01 \pm 1.29$ &$0.82 \pm 0.01$& $\mathbf{31.39 \pm 1.32}$ & $\mathbf{0.84 \pm 0.01}$ \\
    \midrule
    32-Shots &  $31.01 \pm 1.38$ & $0.83 \pm 0.02$ & $\mathbf{32.70 \pm 0.34}$ & $\mathbf{0.87 \pm 0.02}$ & $29.87 \pm 1.54 $ &$0.79 \pm 0.03$ & $\mathbf{31.89 \pm 0.54}$ & $\mathbf{0.85 \pm 0.01}$& $32.20 \pm 1.29$ &$0.86 \pm 0.01$ & $\mathbf{33.12 \pm 0.96}$  & $\mathbf{0.87 \pm 0.01}$ \\
    \bottomrule
  \end{tabular}}
  \vspace{-3mm}
\end{table*}

The learned k-space trajectories using radial trajectory as the initial sampling pattern were demonstrated in Figure~\ref{fig:radial_trj}. Both the initial trajectory and the learned trajectory are presented. The learned trajectories tend to densely sample the central k-space region with higher information density. The learned trajectories are divergent at the peripheral k-space region since the trajectories prefer to acquire appropriate high-frequency components, which are typically sparse in the k-space. 

The radial imaging use straight-line sampling pattern to acquire k-space. In Figure~\ref{fig:radial_trj}, the wavy pattern was formed based on the radial spoke, resulting in a dense sampling of the central k-space. Unlike the radial trajectory, where the sampling density falls off rapidly away from the k-space center, the learned trajectory provides a new strategy to sample the entire k-space with a balanced and uniform sampling density distribution. The influence of the physical constraints on the optimized trajectories is also demonstrated in the middle of Figure~\ref{fig:radial_trj}. In these examples, the trajectories were only optimized on image loss from the first term of Eq.~(\ref{eq:loss}). While the unconstrained trajectories attempted to rapidly explore large sampling space for covering more k-space information, the resulted trajectories tend to traverse with abrupt turns, leading to irregular non-smooth sampling patterns which are difficult to be implemented in MRI scanners. However, the learned trajectories under physical constraints can produce a hardware-friendly waveform for practical implementation.

The point-spread function (PSF) is also demonstrated for each corresponding trajectory in Figure~\ref{fig:radial_trj}. Compared with the fixed PSF, the learned trajectory can lead to a PSF with reduced side lobes and more homogeneous sampling of the neighboring pixels. This can result in reduced structural and aliasing imaging artifacts in the undersampled images.

The context-awareness of the trajectory optimization is also investigated by comparing the optimized trajectories for datasets with different anatomies. Figure~\ref{fig:diff-dataset} illustrates the learned trajectories for brain and knee datasets, respectively, using an initialization of the same Cartesian trajectory. There are differences between the exemplified k-space for the brain and knee due to the difference of the imaged objects. The learned trajectories can realize this feature and correctly characterize the difference of the k-space density distribution. More specifically, the learned trajectory has more fluctuation and coverage for the scattered knee k-space than the more centralized brain k-space.

The reconstructed images from the learned trajectories using radial trajectory were demonstrated in Figure~\ref{fig:POST-ALL}. These images were compared with the images directly reconstructed using fixed trajectory. The framework was slightly modified by removing the trajectory optimization network and only training the end-to-end reconstruction U-Net using the standard supervised learning approach. The qualitative evaluation of brain images proves the improved image reconstruction using our proposed method. As illustrated in Figure~\ref{fig:POST-ALL}, the reconstructed images from learned trajectories are consistently better than those from the fixed trajectories for each type. More specifically, the learned radial trajectories provided improved reconstruction performance compared to their fixed counterparts in Figure~\ref{fig:POST-ALL} for the brain images at the AXT1POST sequence (More results in Appendix). Notably, the intermediate images directly obtained from the RSS reconstruction were shown at the top row of Figure~\ref{fig:POST-ALL} for the learned and fixed trajectories. It is evident that the learned trajectory can better remove structural and aliasing artifacts and provided more realistic image features and accurate image contrast than that of the fixed trajectory at the same level of acceleration, indicating the efficacy of the learning-based trajectory optimization.

\subsubsection{Quantitative Comparison}
The group‐wise quantitative analyses further confirmed the qualitative comparisons as shown in the exemplary figures. Tables~\ref{TableOne} and~\ref{TableTwo} summarized the quantitative comparison between images reconstructed using the fixed and learned Cartesian, radial, and spiral trajectories in all testing AXT1 and AXT1POST brain image datasets at different acceleration levels. In general, there is improved reconstruction quality with the decrease of acceleration level (\textit{i.e.}, more shots) for both fixed and learned trajectories, as indicated by the increased PSNR and SSIM values. At the same acceleration level, the learned trajectories show significantly better reconstruction quality ($p < 0.05$) than the fixed ones in terms of PSNR and SSIM metrics for all sampling patterns on AXT1 and AXT1POST brain images. 

More comparison results can be found in the supplementary document of this paper. We also show its clinical value for segmentation task on BraTS2020~\cite{brats}.

\section{Discussion}

This study demonstrated the feasibility of optimizing MRI k-space acquisition using a deep learning framework consisting of a neural ODE and an image reconstruction network. This framework has achieved a successful joint model for simultaneously learning optimal k-space acquisition and image reconstruction of the raw k-space data. The proposed method was evaluated under Cartesian and Non-Cartesian trajectories to demonstrate the generalization of the method. In several image datasets obtained with different MRI sequences at different anatomical structures, the proposed learning-based k-space optimization can correctly characterize the inherent k-space features. Our method provides an efficient acquisition strategy, meanwhile maintaining high-quality image reconstruction compared to the regular fully sampled Cartesian and Non-Cartesian trajectories, as shown in our results.

A few previous studies have focused on k-space optimization under the compressed sensing framework~\cite{haldar2019oedipus,sherry2020learning,sanchez2020scalable,lazarus2019sparkling}. While most of these methods have shown success in optimizing k-space acquisition on an individual image, it becomes challenging to ensure consistent sampling efficiency and reconstruction performance for a wide variety of image slices. The optimization's performance is also dependent on the assumptions and priors imposed for interpreting the k-space features. Such optimization is primarily heuristic thus could lead to a suboptimal result in the presence of different image features. Recent deep learning-based approaches can be more robust and adaptive to complicated image features. Because the deep learning model can be trained against many image datasets, it learns comprehensive image contents and their corresponding k-space features, leading to more accurate k-space representation. In our study, we applied a neural ODE to characterize the dynamics of k-space acquisition. Unlike several previous deep learning-based approaches, this new approach does not assume the k-space data distribution; instead, it relies solely on the learning process to characterize essential k-space features (Figure~\ref{fig:diff-dataset}). In addition, neural ODE is a generic framework capable of optimizing any k-space trajectories.In the current study, we only demonstrated the optimization for Cartesian, radial, and spiral trajectories. The future extension could be applied for other trajectories such as echo planner imaging~\cite{stehling1991echo}, propeller acquisition~\cite{pipe1999motion}, and other hybrid sampling patterns~\cite{noll1997multishot}. Extension to multi-dimensional data acquisition such as 3D imaging~\cite{johnson2017hybrid} and dynamic imaging~\cite{tamir2017t2} is also possible given sufficient training datasets and computing power. 

In addition, we demonstrated the influence of physical constraints on the k-space optimization. In contrast to the unconstrained optimization, the k-space optimization conditioned on MRI physics could lead to rapid k-space coverage with maximally allowable traverse patterns, resulting in homogeneous image acquisition as indicated by the PSFs in Figure~\ref{fig:radial_trj}. In the current study, we only considered the factors of maximum gradient amplitude and slew rate. In the future study, a more realistic physical model accounting for eddy current effects~\cite{ahn1991analysis} could be imposed to estimate more robust trajectory patterns.

Our framework not only estimates the optimal k-space acquisition but also simultaneously trains a deep learning-based image reconstruction network using the learned k-space data. Inspired by adversarial learning, the joint k-space optimization and image reconstruction can maximize the optimization performance in a competing manner. The mutual competing nature of training can improve both the k-space optimization network and the reconstruction network collaboratively, leading to a better performance than training each network separately. Undoubtedly, the learned k-space trajectory is subject to the selection of a reconstruction network. Unlike the optimized sampling patterns under compressed sensing, where the trajectories typically follow k-space distribution density, the learned trajectories using neural ODE provide unique k-space characterization that could be more suitable for the jointly learned reconstruction network. In the current study, we only applied U-Net for image reconstruction. Many recently proposed network architectures tailored for MRI reconstruction, such as unrolled networks~\cite{hammernik2018learning,biswas2019dynamic}, domain transfer learning networks~\cite{akccakaya2019scan}, and recurrent neural networks~\cite{oh2021k}, could be potential candidates to further improve the image reconstruction performance in combination with the k-space optimization neural ODE in our framework. In addition, the training loss applied in the current study is solely based on the difference between the fully sampled reference and the estimated image. Future work will explore a more comprehensive loss function to incorporate useful image priors and imaging parameters to create a more robust model.

Our proof-of-concept study has several limitations.  Due to the GPU memory limit, we must apply an RSS reconstruction to combine multi-coil images into a single image for image reconstruction. However, using multi-coil k-space data could further improve the reconstruction performance through advanced reconstruction networks. To demonstrate the feasibility of our framework, we only investigated retrospective acquisitions of the existing image datasets. The current study is ongoing to evaluate the learned trajectories on real MRI through implementing specific MRI sequences. Finally, the optimized imaging method has not been evaluated for pathology diagnosis in a large patient cohort to illustrate the clinical value.

\section{Conclusion}

Our study presented a novel deep learning framework to learn MRI k-space trajectory optimization. We have shown that the proposed method consistently outperforms the regular fixed k-space sampling strategy. The optimization is efficient and adaptable for various Cartesian and Non-Cartesian trajectories at different image sequences, contrast, and anatomies. The proposed method provides a new opportunity for improving rapid MRI, ensuring optimal acquisition while maintaining high-quality image reconstruction.

\section*{Acknowledgement}
This work was supported by the Academy of Finland for ICT 2023 project (grant 328115), Academy Professor project EmotionAI (grants 336116, 345122), by Ministry of Education and Culture of Finland for AI forum project, and Infotech Oulu. As well, the authors wish to acknowledge CSC-IT Center for Science, Finland, for computational resources. 

{\small
\bibliographystyle{ieee_fullname}
\bibliography{egbib}
}

\newpage
\appendix
\noindent \textbf{\LARGE{Appendix}}


\section{Implementation details}

\begin{figure*}
    \centering
    \subfloat[Neural ODE]{\includegraphics[width=0.25\textwidth]{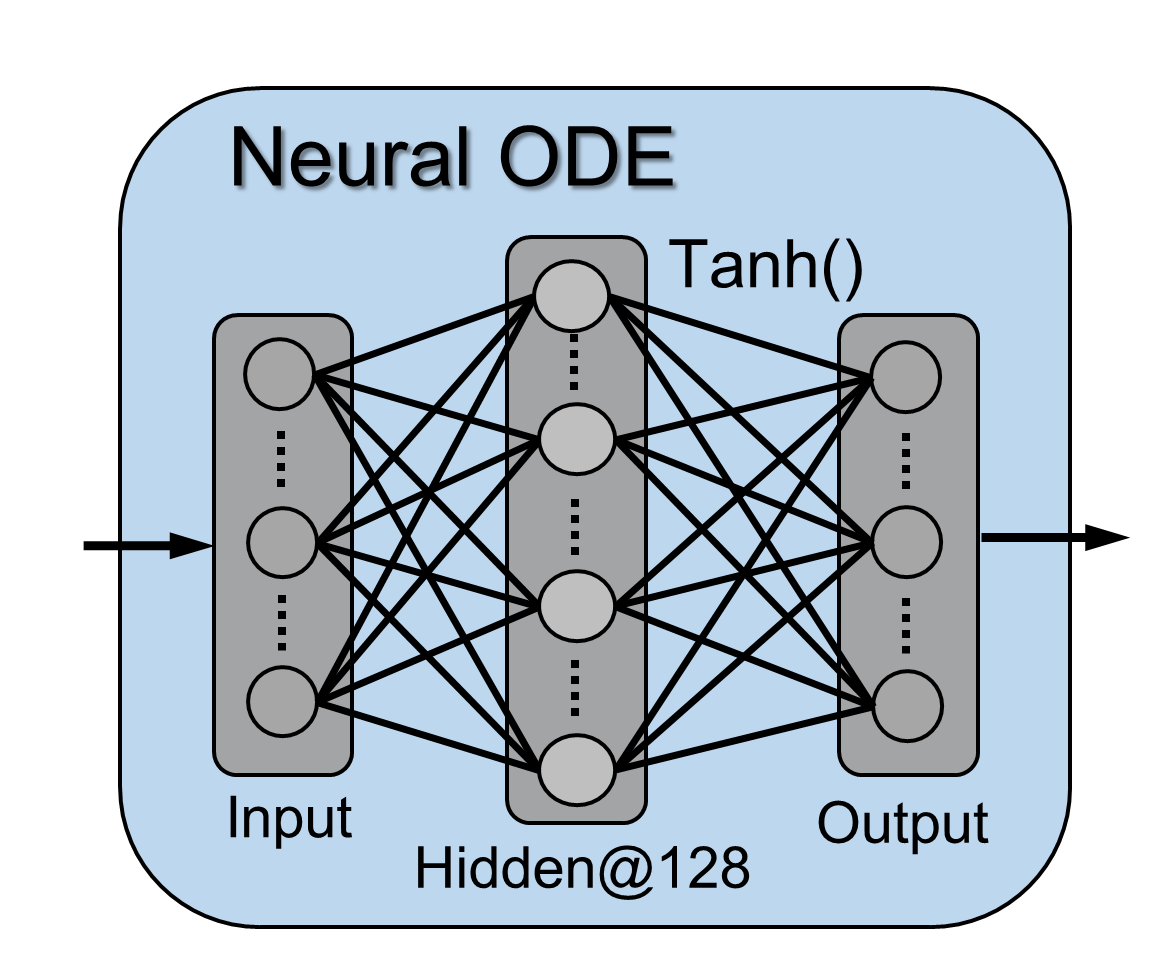}}
    \subfloat[Reconstruction CNN]{\includegraphics[width=0.7\textwidth]{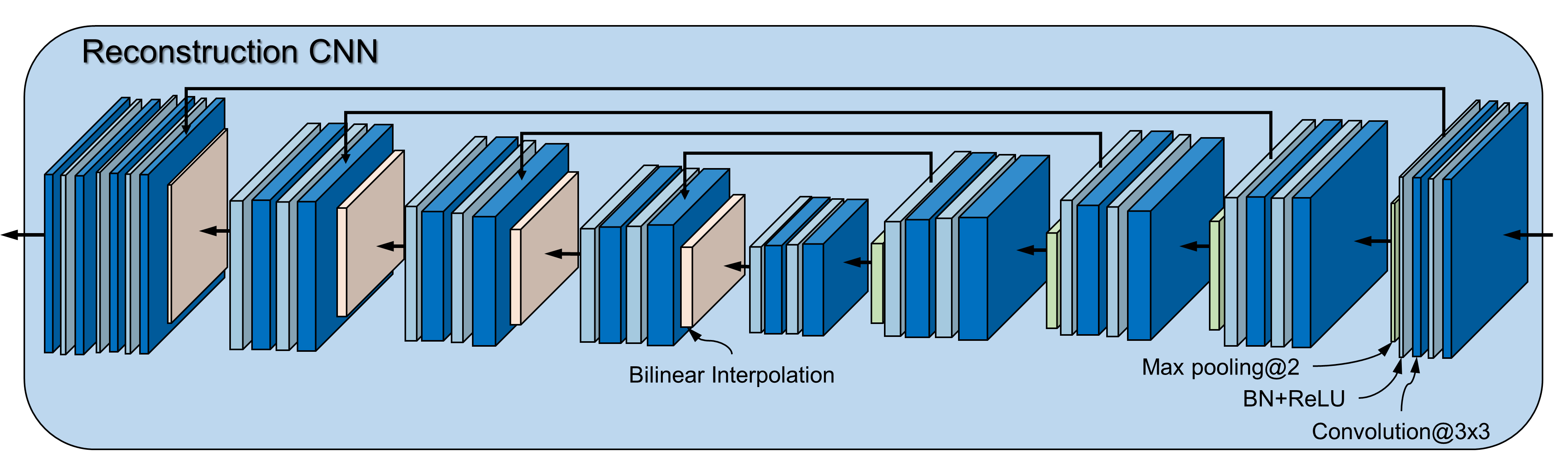}}
    \caption{\footnotesize{Schematic illustration of the proposed framework. \textbf{(a)} Detailed architecture of the neural ODE using a fully connected network. \textbf{(b)} A reconstruction CNN using a U-Net architecture.}}
    \label{fig:frame2}
\end{figure*}

Our proposed models were designed using PyTorch\cite{Paszke2019PyTorchAI} deep learning package. As shown in Figure~\ref{fig:frame2} (a), the neural ODE uses a fully connected network (FCN) consisting of two fully connected layers and a Tanh activation function. The k-space location (\textit{i.e.}, [kx, ky] in 2D acquisition) of the sampling points were concatenated together for all shots and used as the input of the FCN. The output size kept the same as the input. As shown in Figure~\ref{fig:frame2} (b), like many medical image reconstruction studies, a U‐Net~\cite{ronneberger2015u} was used as the end-to-end reconstruction network to remove the residual artifacts and noises in the intermediate images, obtained using nuFFT and adjoint nuFFT operations on the optimized k-space trajectories. The U‐Net structure comprises an encoder network with four layers of the down-sampling channel and a decoder network with a mirrored and reversed encoder structure. Multiple skip connections are used to concatenate entire feature maps from encoder to decoder to enhance mapping performance. In addition, because of the GPU memory limit, the multi-channel images were combined into one image using a root-sum-of-squares reconstruction (RSS)~\cite{roemer1990nmr} and then used as the input of the U-Net network. 

\section{More Results}
We provide qualitative comparisons with PILOT~\cite{weiss2019pilot} under the same protocol. Here, we only use 1000 sampling points for each spoke. As shown in Table~\ref{TableOne}, our method can be much better especially when acceleration is high. As PILOT optimizes the acquisition by directly introducing a learnable matrix, which is surely sensitive to initialization and would easily stuck in suboptimal points.  

\begin{table}[h]
  \caption{ \small{Comparisons on brain AXT1 images, using radial.}}
  \label{TableOne}
  \centering
  \scalebox{0.76}{\begin{tabular}{lcccc}
    \hline
    \multirow{2}{*}{Acc. Level} &\multicolumn{2}{c}{PILOT} & \multicolumn{2}{c}{Ours}     \\
    &PSNR &SSIM &PSNR &SSIM  \\
   \hline  
 16-shots &$28.62 \pm 0.66 $ & $0.77 \pm 0.01$ & $\mathbf{31.03 \pm 0.59 }$ & $\mathbf{0.82 \pm 0.01}$ \\
    \hline
    32-shots  &$31.24 \pm 0.61 $ & $0.83 \pm 0.01$ & $\mathbf{33.06 \pm 1.23 }$ & $\mathbf{0.86 \pm 0.01}$ \\
    \hline
  \end{tabular}}
\end{table}

To demonstrate the capability of our proposed method for optimizing both Cartesian and Non-Cartesian trajectories, we investigated initializing the trajectory optimization using three different k-space sampling patterns, including Cartesian, radial, and spiral, respectively. For the Cartesian initialization, a 10\% central k-space was fully sampled, and N lines were sampled in the rest of the k-space with equal distance between any two adjacent encoding lines. For the spiral initialization, a uniform density spiral interleave was first created, and the rest $N-1$ interleaves were created by linearly rotating $2\pi/N$ degrees.

\begin{figure}[h]
    \centering
    \includegraphics[width = 0.4 \textwidth]{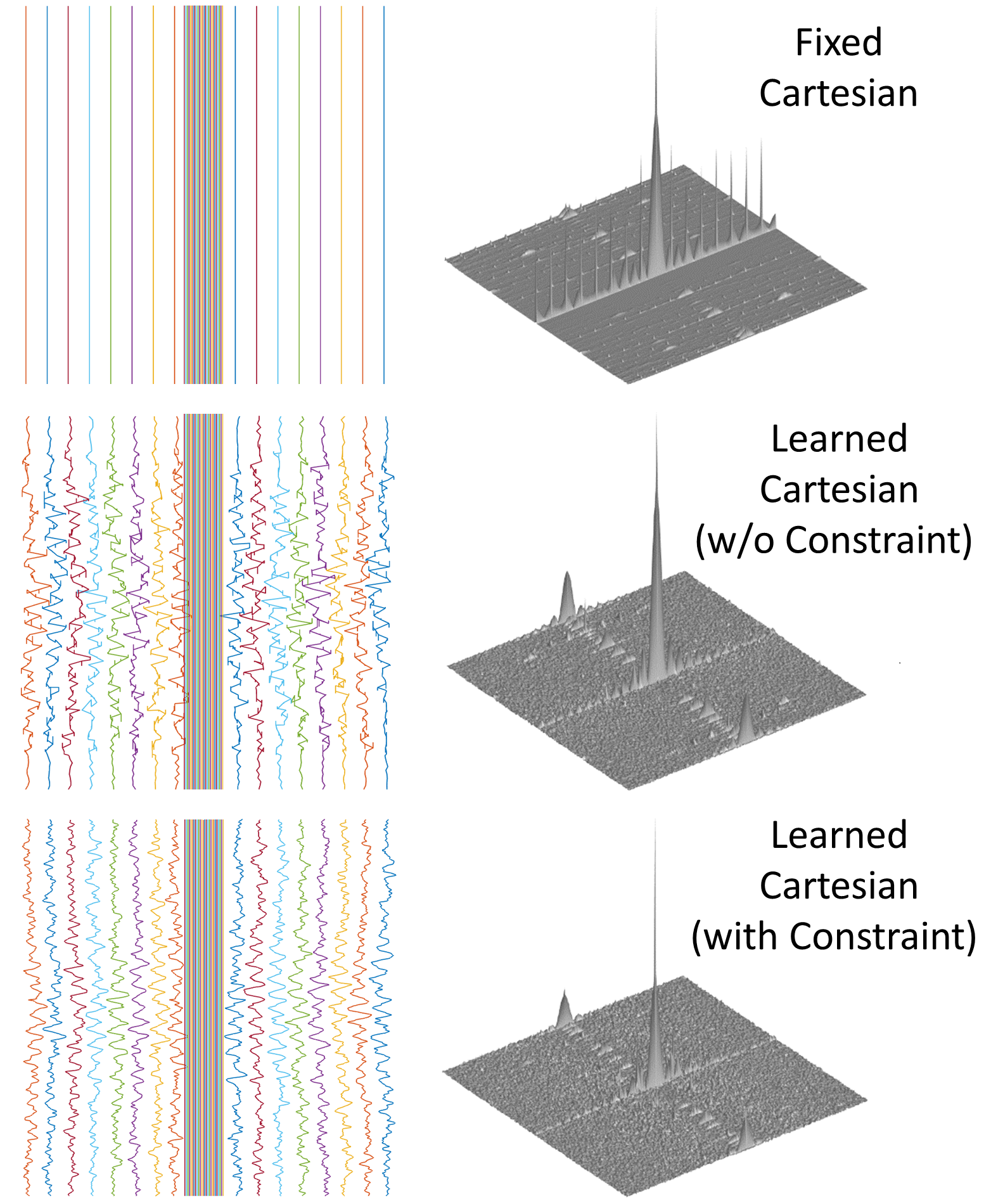}
    \caption{\footnotesize{Trajectory comparison between fixed and learned Cartesian trajectory with acceleration factor AF = 6 (16 phase encoding lines) optimized on the knee datasets. \textbf{Top}: initial Cartesian trajectory. \textbf{Middle}: learned Cartesian trajectory without physical constraints. \textbf{Bottom}: learned Cartesian trajectory with physical constraints. The corresponding PSF is shown for each trajectory.}}
    \label{fig:cartesian_trj}
\end{figure}

\begin{figure}[h]
    \centering
    \includegraphics[width = 0.38 \textwidth]{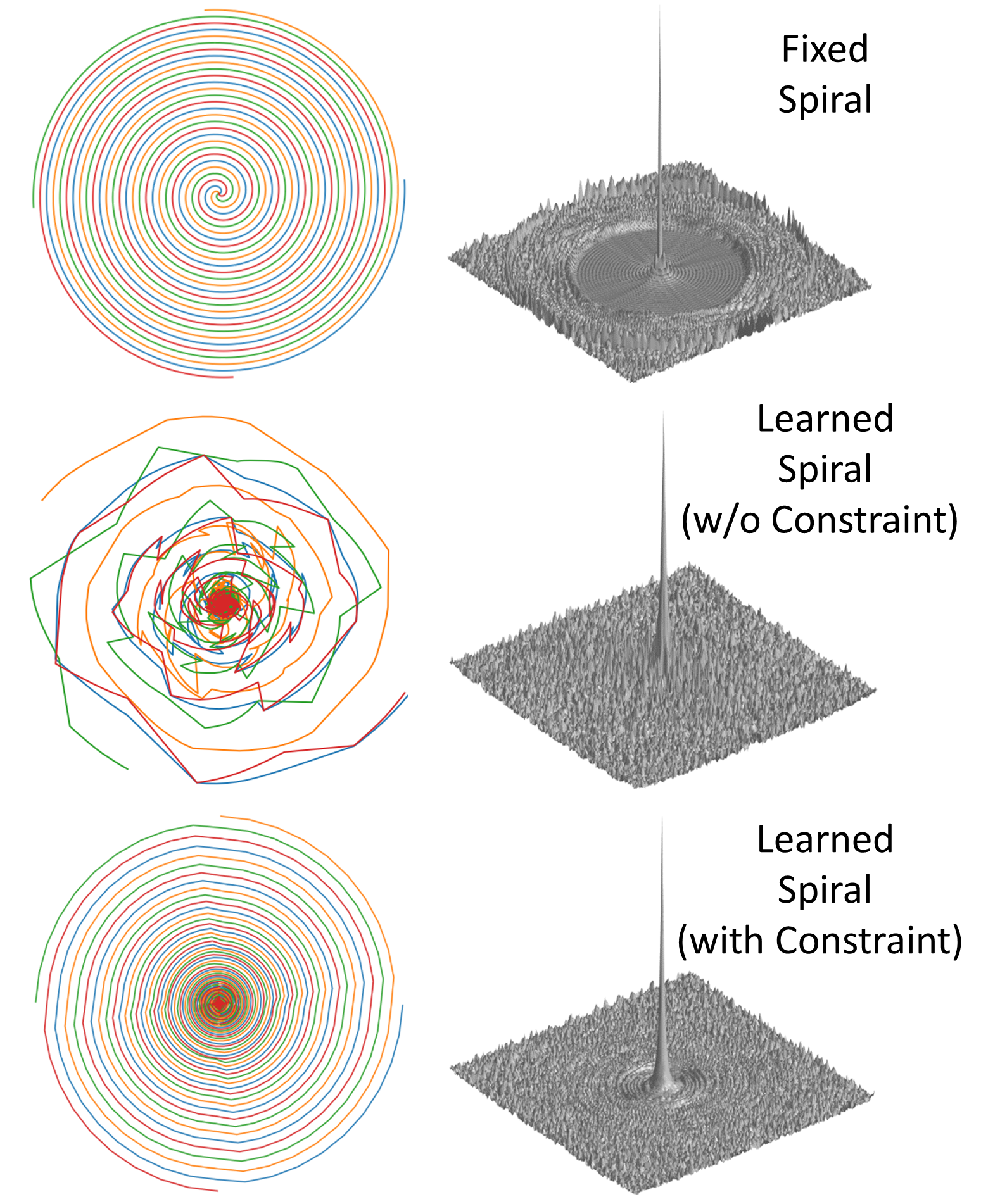}
    \caption{\footnotesize{Trajectory comparison between fixed and learned spiral trajectory with 4 interleaves, optimized on the brain AXT1 datasets. \textbf{Top}: initial spiral trajectory. \textbf{Middle}: learned spiral trajectory without physical constraints. \textbf{Bottom}: learned spiral trajectory with physical constraints. The corresponding PSF is shown for each trajectory.}}
    \label{fig:spiral_trj}
\end{figure}

The learned k-space trajectories using Cartesian, and spiral trajectory as the initial sampling pattern were demonstrated in Figures~\ref{fig:cartesian_trj}, and~\ref{fig:spiral_trj}, respectively. The Cartesian imaging uses straight-line sampling pattern to acquire k-space. In Figure~\ref{fig:cartesian_trj}, the learned trajectory uses a Cartesian trajectory as an initial sampling pattern. A wavy sampling pattern was formed at the canter region of each phase encoding line, becoming more efficient in acquiring the high-density k-space region than a straight line. Spiral imaging uses a curved sampling pattern to acquire k-space. In Figure~\ref{fig:spiral_trj}, the spiral trajectory was further optimized to cover the k-space more efficiently. The initial uniform density spiral spoke adaptively concentrated into the central k-space region with a slightly wavy pattern, which assembles the variable density spiral sampling, which was previously shown to be more efficient in spiral imaging~\cite{tsai2000reduced}.

The influence of the physical constraints on the optimized trajectories is also demonstrated in the middle of Figures~\ref{fig:cartesian_trj}, and~\ref{fig:spiral_trj}.  While the unconstrained trajectories attempted to rapidly explore large sampling space for covering more k-space information, the resulted trajectories tend to traverse with abrupt turns, leading to irregular non-smooth sampling patterns which are difficult to be implemented in MRI scanners. However, the learned trajectories under physical constraints can produce a hardware-friendly waveform for practical implementation.

The point-spread function (PSF) is also demonstrated for each corresponding trajectory in Figures~\ref{fig:cartesian_trj}, and~\ref{fig:spiral_trj}. Compared with the fixed PSF, the learned trajectory can lead to a PSF with reduced side lobes and more homogeneous sampling of the neighboring pixels. This can result in reduced structural and aliasing imaging artifacts in the undersampled images.


\begin{figure*}[h]
    \includegraphics[width = 0.99\textwidth]{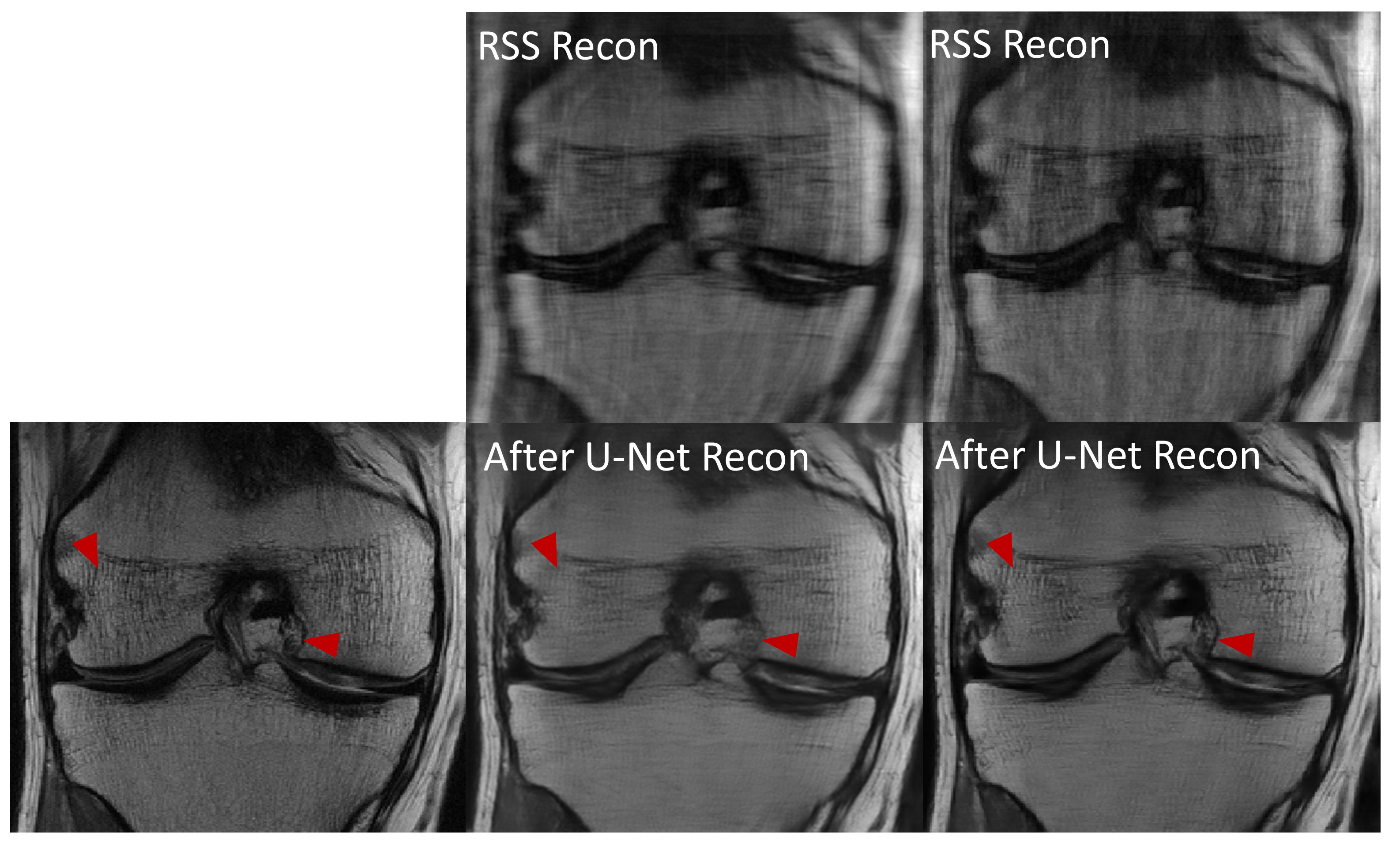}
    \caption{\footnotesize{MRI reconstruction comparison between fixed and learned Cartesian trajectory at acceleration rate AF = 4.4 (32 phase encoding lines) for knee images. \textbf{Left column}: ground truth fully sampled image. \textbf{Middle column}: upper is the input image getting from RSS with fixed trajectory. Bottom is the reconstructed image from U-Net. \textbf{Right column}: upper is the input image getting from RSS with learned trajectory. Bottom is the reconstructed image from U-Net. The learned trajectory provides more realistic image feature recovery and sharper image quality than the fixed trajectory, as indicated by the red arrows.}}
    \label{fig:Knee-All}
\end{figure*}

\begin{figure*}[!htb]
    \includegraphics[width = 0.99\textwidth]{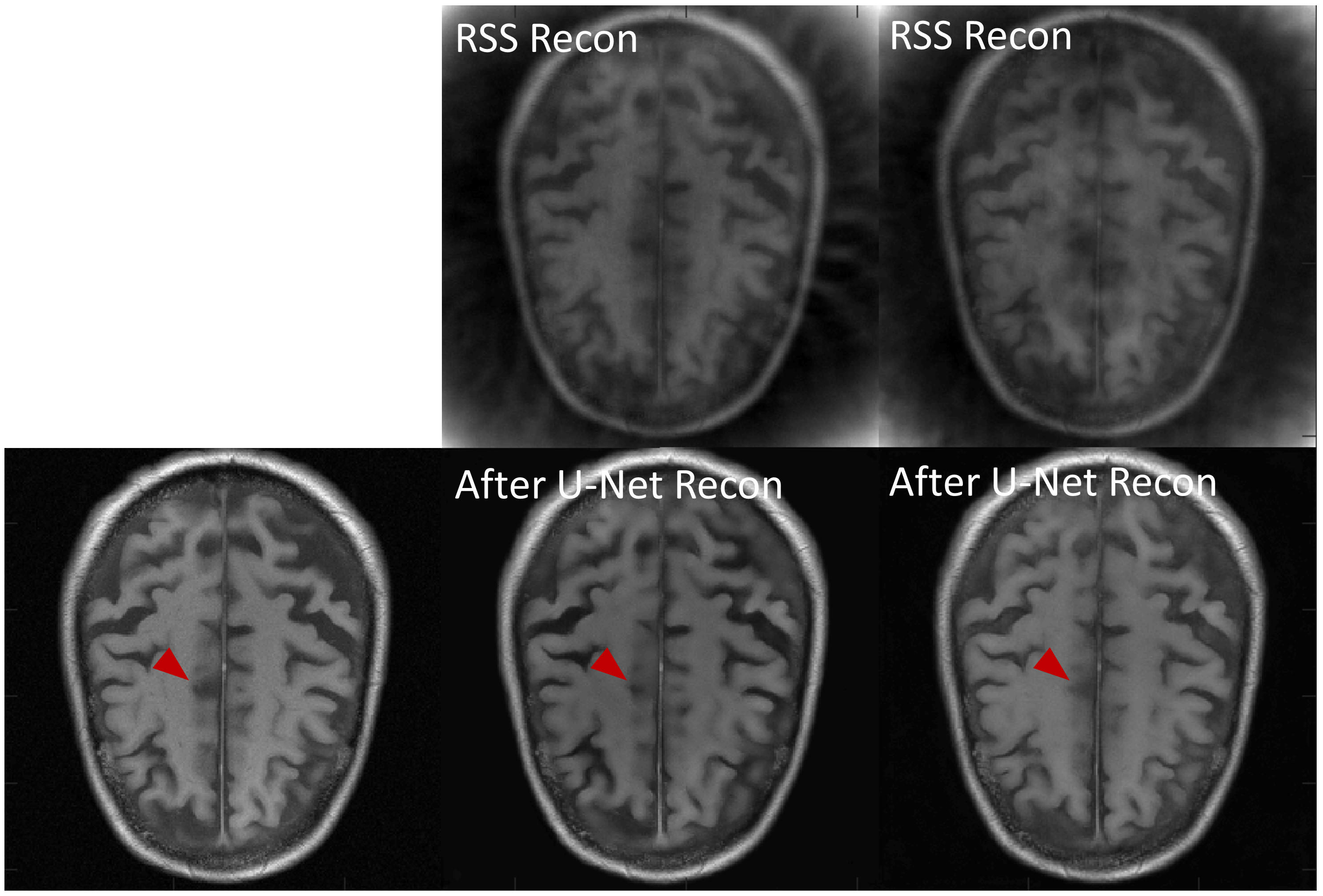}
    \vspace*{0pt}
    \caption{\footnotesize{MRI reconstruction comparison between fixed and learned spiral trajectory with 16 interleaves for brain AXT1 images. \textbf{Left column}: ground truth fully sampled image. \textbf{Middle column}: upper is the input image getting from RSS with fixed trajectory. Bottom is the reconstructed image from U-Net. \textbf{Right column}: upper is the input image getting from RSS with learned trajectory. Bottom is the reconstructed image from U-Net. The learned trajectory provides sharper images and more image details than the fixed trajectory indicated by the red arrows.}} \label{fig:PRE-ALL}
\end{figure*}

The reconstructed images from the learned trajectories using Cartesian, and spiral trajectories were demonstrated in Figures~\ref{fig:Knee-All}, and~\ref{fig:PRE-ALL}, respectively. These images were compared with the images directly reconstructed using fixed trajectory. The framework was slightly modified by removing the trajectory optimization network and only training the end-to-end reconstruction U-Net using the standard supervised learning approach. The qualitative evaluation of knee and brain images proves the improved image reconstruction using our proposed method. As illustrated in Figures~\ref{fig:Knee-All}, and~\ref{fig:PRE-ALL}, the reconstructed images from learned trajectories are consistently better than those from the fixed trajectories for each type. More specifically, Figure~\ref{fig:Knee-All} provides an example of a reconstructed knee image using a learned Cartesian trajectory at an acceleration factor (AF) of 4.4. This figure shows that the learned trajectory provides better image features, improved image sharpness, and more detail recovery due to their optimized k-space coverage. The learned Cartesian trajectory outperformed the regular Cartesian trajectory at the same acceleration rate. Likewise, the learned spiral trajectories provided improved reconstruction performance compared to their fixed counterparts in Figure~\ref{fig:PRE-ALL} for the brain images at the AXT1 sequence. Notably, the intermediate images directly obtained from the RSS reconstruction were shown at the top row of Figures~\ref{fig:Knee-All}, and~\ref{fig:PRE-ALL} for the learned and fixed trajectories. It is evident that the learned trajectory can better remove structural and aliasing artifacts and provided more realistic image features and accurate image contrast than that of the fixed trajectory at the same level of acceleration, indicating the efficacy of the learning-based trajectory optimization.

\begin{figure*}[!htb]
\centering
    \includegraphics[width = 0.93\textwidth]{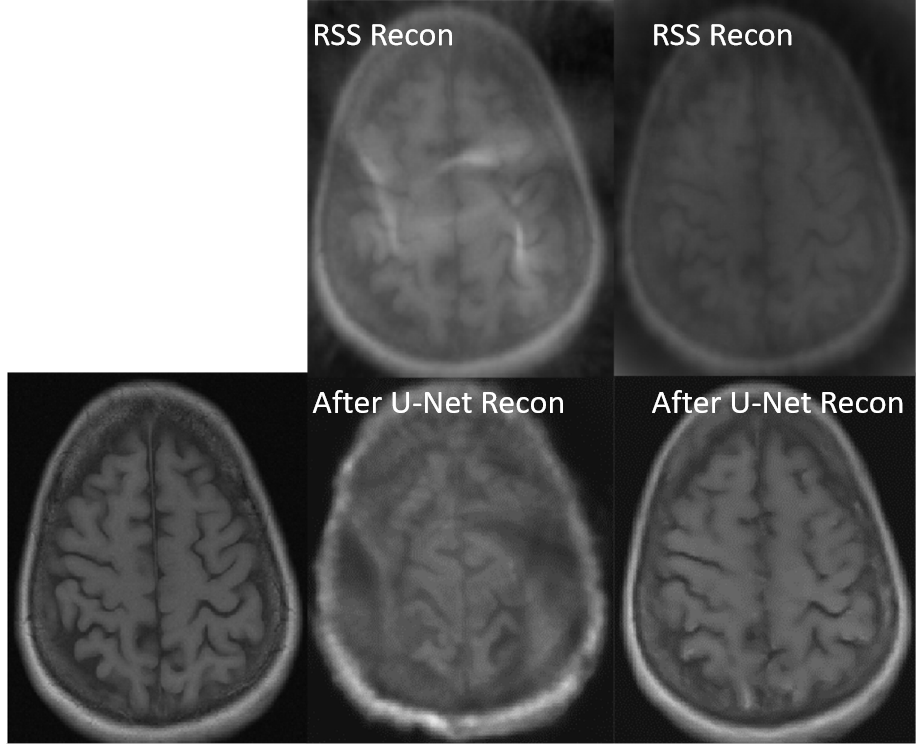}
    \vspace*{0pt}
    \caption{\footnotesize{MRI reconstruction comparison between fixed and learned spiral trajectory with 4 interleaves for brain AXT1 images. \textbf{Left column}: ground truth fully sampled image. \textbf{Middle column}: upper is the input image getting from RSS with fixed trajectory. Bottom is the reconstructed image from U-Net. \textbf{Right column}: upper is the input image getting from RSS with learned trajectory. Bottom is the reconstructed image from U-Net. The learned trajectory provides sharper images and more image details than the fixed trajectory.}} \label{fig:PRE4}
\end{figure*}

We also explored the generalization ability of the proposed method to high acceleration level. Here, we undersampled the k-space data using only four spiral interleaves. As illustrated in Figure~\ref{fig:PRE4}, the learned trajectory provides much sharper images and more image details than the fixed trajectory, of which the provided MR images are far away from satisfaction as the artifacts are very everywhere.

The context-awareness of the trajectory optimization is also investigated by comparing the optimized trajectories for datasets with different anatomies. Figure~\ref{fig:diff-dataset} illustrates the learned trajectories for brain and knee datasets, respectively, using an initialization of the same Cartesian trajectory. There are differences between the exemplified k-space for the brain and knee due to the difference of the imaged objects. The learned trajectories can realize this feature and correctly characterize the difference of the k-space density distribution. More specifically, the learned trajectory has more fluctuation and coverage for the scattered knee k-space than the more centralized brain k-space.

\section{Segmentation on undersampled MRI}
\begin{figure*}
  \centering
  \includegraphics[width=0.95\linewidth]{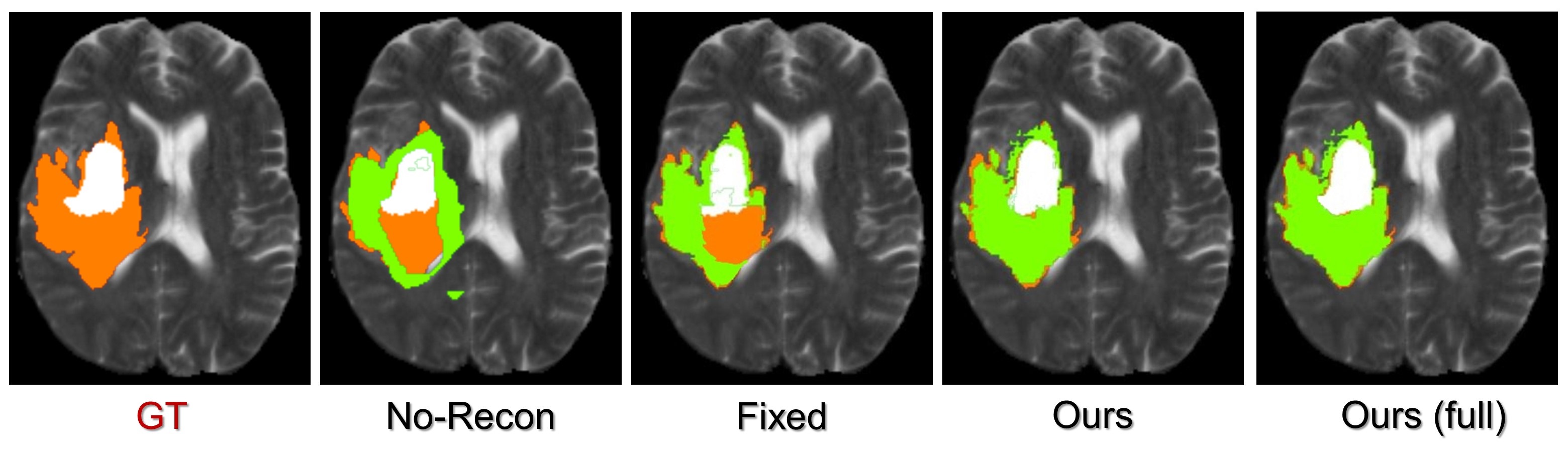}
  \caption{\small{Tumor segmentation on BraTS2020~\cite{brats} dataset. The MRI is accelerated by 8-fold. Our method can provide a comparable segmentation result when compared to method using fully sampled MRI.}}
   \label{fig:seg}
\end{figure*}

To further assess the clinical value of the proposed method, we apply it to the tumor segmentation task on undersampled MRI from the BraTS2020~\cite{brats} datatset. Here, the MRI is accelerated for 8-fold using radial trajectory. As illustrated in Fig.~\ref{fig:seg}, the segmentation model will collapse when directly training  from the undersampled MRI (No-Recon). Using the proposed method, we can get a similar performance when compared to the model based on the fully sampled data. Our method is much more efficient as we accelerate it for eight times.   

\end{document}